\documentclass[a4paper, superscriptaddress, amsfonts, amssymb, amsmath, reprint, showkeys, nofootinbib, twoside]{revtex4-1}
\usepackage[english]{babel}
\usepackage[utf8]{inputenc}
\usepackage{xcolor}
\usepackage{bm}
\usepackage{graphicx}
\usepackage{float}
\usepackage{amsmath}
\usepackage{amsfonts,amssymb,amsthm}
\usepackage[colorinlistoftodos, color=green!40, prependcaption]{todonotes}
\usepackage{amsthm}
\usepackage{mathtools}
\usepackage{physics}
\usepackage{xcolor}
\usepackage{graphicx}
\usepackage[left=23mm,right=13mm,top=35mm,columnsep=15pt]{geometry} 
\usepackage{adjustbox}
\usepackage{placeins}
\usepackage[T1]{fontenc}
\usepackage{lipsum}
\usepackage{csquotes}
\usepackage[utf8]{inputenc}
\usepackage[pdftex, pdftitle={Article}, pdfauthor={Author}]{hyperref} 

\begin{document}
\title{Redundant and synergistic interactions in a complex network of single-transistor electronic chaotic oscillators and in neurophysiological recordings}

\author{Chiara Barà}
    \affiliation{Institute of Intelligent Industrial Technologies and Systems for Advanced Manufacturing, National Research Council, Milan 20133, Italy}
    \affiliation{Department of Engineering, University of Palermo, 90128 Palermo, Italy}
\author{Yuri Antonacci}
    \affiliation{Department of Engineering, University of Palermo, 90128 Palermo, Italy}
\author{Laura Sparacino}
    \affiliation{Department of Engineering, University of Palermo, 90128 Palermo, Italy}
\author{Ariosky Areces Gonzalez}
    \affiliation{School of Life Science and Technology, University of Electronic Science and Technology of China, 611731 Chengdu, China}
    \affiliation{School of Technical Sciences, University of Pinar del Río “Hermanos Saiz Montes de Oca”, Pinar del Rio, Cuba}
\author{Manyu Zhao}
    \affiliation{School of Life Science and Technology, University of Electronic Science and Technology of China, 611731 Chengdu, China}
\author{Longxiang Fu}
    \affiliation{School of Life Science and Technology, University of Electronic Science and Technology of China, 611731 Chengdu, China}
\author{Pedro A. Valdes-Sosa}
    \affiliation{School of Life Science and Technology, University of Electronic Science and Technology of China, 611731 Chengdu, China}
    \affiliation{Department of Neuroinformatics, Cuban Neuroscience Center, 11600 Havana, Cuba}
\author{Hiroyuki Ito}
    \affiliation{Nano Sensing Research Unit, Institute of Innovative Research, Institute of Science Tokyo, 226-8503 Yokohama, Japan}
\author{Mattia Frasca}
    \affiliation{Department of Electrical Electronic and Computer Engineering, University of Catania, 95131 Catania, Italy}
\author{Luca Faes}
    \affiliation{Department of Engineering, University of Palermo, 90128 Palermo, Italy}
    \affiliation{Faculty of Technical Sciences, University of Novi Sad, Novi Sad 21102, Serbia}
\author{Ludovico Minati}
    \affiliation{School of Life Science and Technology, University of Electronic Science and Technology of China, 611731 Chengdu, China}
    \affiliation{Nano Sensing Research Unit, Institute of Innovative Research, Institute of Science Tokyo, 226-8503 Yokohama, Japan}
    \affiliation{Center for Mind/Brain Sciences (CIMeC), University of Trento, 38123 Trento, Italy}
    \email{lminati@ieee.org,lminati@uestc.edu.cn}

\begin{abstract}
Complex networks often exhibit emergent behaviors, where simple dyadic interactions yield collective dynamics that cannot be explained by examining the system's units individually or in pairs. Understanding how redundant and synergistic interaction emerges from elementary connectivity patterns is important in characterizing the behavior of physical, biological, and engineering systems. In this study, the information-theoretic framework of Partial Information Decomposition (PID) is employed to investigate how pairs of signals measured at the nodes of large network systems contribute individually and in cooperation with each other to determine the overall state of the network. The analyzed systems are networks of numerically simulated R\"ossler oscillators and physical single-transistor electronic chaotic oscillators, reproducing purely pairwise and symmetric links in biological neuronal cultures, and cortical electroencephalographic networks assessed in humans. In the two settings, PID was extensively applied to decompose the information brought by pairs of signals to the overall state of the system, assessed respectively as the dynamic regime (from asynchronous chaos to synchronization) and the subject condition (open vs. closed eyes). Our approach highlights the coexistence of redundant and synergistic interplays in determining the effect of the pairwise dynamics on the system’s state. Specifically, we demonstrate how, in a highly synchronized system, where units act following overlapped redundant behaviors, their joint consideration helps determine the system's state though their synergistic interactions. Our findings reveal the emergence of non-trivial behaviors in networked system based on pairwise connections and straightforward neural states, highlighting the need for using appropriate analytical tools to capture complex phenomena.
\end{abstract}

\maketitle

\footnotesize
When oscillators in electronic circuits or cortical areas in the brain interact with each other, as a network, they may result in complex behaviors that cannot be explained by observing and describing the units individually. Pairs of units can redundantly and synergistically interact to determine the behavior of the overall network system, resulting in their overlapping and emerging contributions. We follow an unconventional approach in studying these high-order interactions: departing from the traditional approach where all the analyzed variables are measured at the network nodes, we assume that the target variable is the discrete state of the system. With this novel perspective, our findings on systems of different nature, spanning from simulated networks of chaotic oscillators to brain networks, reveal how complex behaviors naturally emerge from system configurations characterized by very elementary coupling or simple functionality. Beyond these results, the relevance of our approach extends to the general problem of relating high-order mechanisms and behaviors in network systems.

\normalsize

\section{Introduction}

Complex systems composed of many interacting units often collectively exhibit behaviors that cannot be explained exclusively in terms of the sum of their pairwise correlations. Even networks endowed with only dyadic structural connections often self-organize to generate configuration patterns that exceed the structural connections, resulting in the establishment of nonlinear, high-order interactions that cannot be factorized \cite{arshinov2003causality, boccaletti2006complex, ladyman2013complex, battiston2021physics, rosas2022disentangling}. Such phenomena are evident across a wide range of real-world applications, from social and behavioral sciences through nonlinear physics and neuroscience \cite{barabasi2013network}. For instance, in a network of coupled oscillators, various synchronization regimes emerge, including clustering and remote synchronization \cite{ponce2015resting, schaub2016graph}, as well as chimera states, where some units synchronize while others remain desynchronized \cite{abrams2004chimera}. These phenomena have been similarly observed in brain dynamics \cite{minati2015synchronization,minati2024chaotic, minati2025spontaneous}: neuronal populations synchronize within a local cluster or across distant regions without structural connections during the execution of cognitive function \cite{belykh2008cluster, vlasov2017hub}, or show a chimera-like state dividing the brain into regions with synchronous and asynchronous oscillatory dynamics under some pathological (e.g., in epilepsy distinguishing between normal and seizure brain area activities \cite{andrzejak2016all}) or functional (e.g., during cognitive multitasking activities where distinct cortical areas synchronizing only within specific task-related networks \cite{calim2020chimera}) conditions. Moreover, as brain activity continuously shifts across different metastable states, it exhibits a dynamic behavior that resembles that of certain chaotic attractors \cite{babloyantz1996brain, mateos2019impact}. The existence of non-trivial, high-order interactions underlying the complex behavior of brain circuits is emphasized by these observations. Accordingly, an optimal balance of functional segregation and integration has been demonstrated to facilitate specialized activity and global synchrony, thus ensuring the typical flexibility of a healthy brain \cite{tononi1994measure, bullmore2009complex, sporns2004organization, patton2025changing}.

Conventionally, analysis of brain connectivity and network dynamics has focused on pairwise interactions among the units of the systems under investigation. This has been achieved through the use of simple correlation or coherence metrics between pairs of signals that describe the dynamics of the units' system, or through graph-theoretic approaches \cite{van2010graph, craddock2015connectomics, chiarion2023connectivity}. However, to characterize the non-trivial interplays within these complex systems, measures are needed that can capture higher-order interactions, where a group of units may carry information that no single pair does \cite{battiston2021physics, rosas2022disentangling}. In this context, the information-theoretic framework provides notions that extend beyond the bivariate connectivity context, as seen in the metrics of high-order functional connectivity \cite{herzog2022genuine} and in the decomposition of the information transfer \cite{faes2016information} or the information rate \cite{faes2022new} among multiple signals, which have been introduced to quantify the interplay among three or more units in a network. In particular, the recently introduced notions of redundancy and synergy provide a principled perspective for characterizing the balance between segregation and integration in terms of how information is distributed across units of a complex system \cite{luppi2024information}. Redundant information occurs when multiple units of a system convey the same knowledge, thereby duplicating the information they share. In contrast, synergy is jointly provided by multiple units and is not available in any single unit, corresponding to emergent information. These concepts are at the core of the analysis of higher-order interactions in both physical systems \cite{haken1977synergetics, zhang2018nonlinear, antonacci2021measuring, hayashibe2022synergetic} and neural networks \cite{gatica2021high, luppi2024information, pope2025time}. In the brain, the duality between functional concepts of synergy and redundancy can be related to the duality between structural concepts of integration and segregation \cite{bassett2017network, varley2023multivariate}. This idea aligns with the established hypothesis that brain function requires a balance of integration and segregation to optimise synergistic interactions between regions and to preserve robust, redundant pathways \cite{luppi2022synergistic, varley2023multivariate}. In this context, the O-Information and O-Information Rate metrics have been introduced to assess the overall balance between redundant and synergistic interplay among multiple units, indicating whether a system is dominated by redundant or synergistic behavior \cite{rosas2019quantifying, faes2022framework, scagliarini2024gradients}. Nevertheless, despite the established efficacy of these metrics  \cite{pirovano2023rehabilitation, kumar2024changes, antonacci2024spectral}, they provide a net balance quantification by collapsing the two complementary aspects of synergy and redundancy into a single metric. This aggregation, however, obscures situations where redundancy and synergy coexist.

To disentangle redundant and synergistic high-order statistical interactions, the Partial Information Decomposition (PID) framework, introduced by Williams and Beer, is recognized as a handy information-theoretical tool \cite{williams2010nonnegative}. This approach decomposes the information that a set of source variables shares with a target, thus their joint mutual information (MI), into the exclusive contribution of each of the sources to the target (unique information), the contribution that the sources shared equally (redundant information), and the one that emerges only when the two sources are considered together and not separately (synergistic information). Considering these different contributions, PID emerges as a powerful tool for providing insights into complex brain dynamics by investigating the interactions between different brain network units that extend beyond neural pairs or a couple of brain regions \cite{wibral2017quantifying, sherrill2021partial, varley2023partial, luppi2022synergistic}. 

Similarly to other metrics of high-order behavior, PID is commonly applied over multiplets of variables measured at the nodes of the analyzed network system to extract measures of unique, redundant and synergistic information which are taken as descriptive of the system state. This standard approach only provides an indirect characterization of the state of the complex system, whose variations are typically assessed by applying statistical tests or machine learning tools on the distributions of the high-order metrics quantified in different states. Moreover, the fact that the number of PID information terms grows super-exponentially with the number of nodes analyzed poses serious problems in the representation, interpretation and empirical computation of the high-order interaction metrics over several variables, making unfeasible to consider all the nodes of large network systems simultaneously \cite{rosas2020reconciling, faes2025predictive}. To overcome these limitations, we follow in this work a different approach that relates the redundant and synergistic modes of information exchange in the network directly to the state of the system, also allowing for a complete yet compact description of such modes within a large system. Specifically, we propose a so-called state-PID (sPID) framework to quantify the unique, redundant, and synergistic contributions to the overall state of the analyzed system, considered as discrete target variable, of the continuous-valued state of two of its units, considered as sources. Extending the approach to any pair of system units taken as sources allows to provide, in the compact form of a connectivity matrix, measures dissecting the higher-order contributions of the units to the behavior of the network. The computationally reliable implementation of the framework is obtained leveraging a recently proposed estimator of the PID for discrete target and continuous source random variables \cite{bara2025partial}. 

The sPID framework is applied to complex networks drawn from three different domains: (i) a numerically simulated network of coupled R\"ossler oscillators; (ii) a physical network of chaotic transistor-based electronic oscillators; and (iii) the human brain network, as described by cortical data. In this systems, sPID quantifies the unique, redundant, and synergistic information that each pair of units carries about a discrete target variable of interest. Here, units are defined as pairs of oscillators in the simulated and hardware systems, and pairs of cortical areas in the brain network, while the target variable is an indicator of the global state of the investigated system. Specifically, in chaotic oscillator systems (both simulated and experimental), the target variable indicates the coupling strength and dynamical regime of the oscillators, ranging from asynchronous chaos to synchronization. In contrast, in the brain network, it indicates the experimental condition, specifically, the resting state with eyes open or closed. This approach provides insight into how pairwise interactions can reflect and influence emergent collective behavior, as well as how redundant and synergistic interactions can coexist within complex systems. Furthermore, our comparative approach illustrates the broad applicability of PID when investigating diverse complex network systems by revealing similarities between neural dynamics and nonlinear chaotic theories.

\section{Methods}\label{method}

Let us consider a network system $\mathcal{X}$ composed of $M$ units, each represented by a random variable $X_m$, with $m = 1, \dots, M$. The collective behavior of these units determines the overall state of the system, which is described by a discrete random variable $Y$ taking values in the set $A_Y=\{1,\dots, Q\}$. Looking at the discrete variable $Y$ as the target variable and at two generic variables $X_i$ and $X_j$ as the source variables, the PID framework can be employed to disentangle the joint MI between the target and the pair of sources -- thus their correlation -- into non-negative information terms indicative of the unique, redundant, and synergistic contributions of the sources to the target variable. This approach can be extended to all possible pairs of source variables, allowing for a comprehensive characterization of the network system's behavior.

This approach, which we denote as state-PID (sPID) to highlight that it investigates the high-order effects of two units and their pairwise correlation on the system state, offers a novel perspective on the already compelling framework of PID, enabling the distinction between the complex modes of interaction of two variables measured from the system at hand in determining a latent variable that classifies the system behavior. In addition to the already recognized advantages of PID, such as the ability to separately quantify the synergistic and redundant contributions that may be obscured by a balanced measure, this approach enables the specific quantification of these contributions for each possible value assumed by the descriptive target variable, facilitating the investigation of how the system observables affect an individual state of the system, and of how these distinct modes of interaction affect the overall state of the system.

\begin{figure*}
    \centering
    \includegraphics{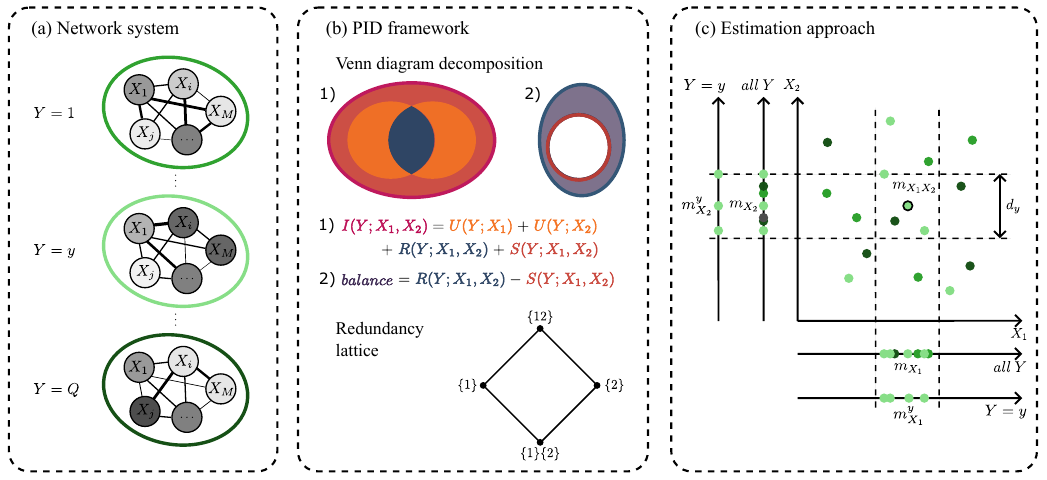}
    \caption{Schematic representation of the pipeline of analysis. In panel ($a$), the network system composed of $M$ units is depicted in multiple configurations dependent on different units' characteristics (represented by varying shades of gray) and connections (represented by lines of varying thickness) collectively determining the state of the overall system's states indicated by the outcomes of the random variable $Y$ depicted by the color of the external circle. In panel ($b$), the state-PID framework applied to decompose the information shared by the target variable $Y$ and two source variables representative of two units of the network, e.g., $X_1$ and $X_2$, is graphically represented through Venn diagrams and the redundancy lattice. Specifically, the Venn diagram representing the decomposition of the joint MI $I(Y;X_1,X_2)$ is reported considering the unique contributions $U(Y;X_1)$ and $U(Y;X_2)$ in orange, the redundant contribution $R(Y;X_1,X_2)$ in blue, and the synergistic contribution $S(Y;X_1,X_2)$ in red. Moreover, the diagram representing the net balance between the redundant and synergistic contributions is shown with the purple area, highlighting how this quantity obscures the effective contribution of the two terms and can assume both positive and negative values, depending on whether redundancy or synergy prevails. The redundancy lattice representing the ordering of the atoms in which the joint MI can be decomposed is also reported. In panel ($c$), a schematic representation of the nearest-neighbor estimation approach used to compute the specific MIs in Eqs. \ref{specMI_1} and \ref{specMI_2} is reported considering a generic sample $\{x_1,x_2\}$ associated to the specific outcome $y$ of the variable $Y$. Part of the figure adapted from Ref. [\citenum{bara2025partial}].}
    \label{fig_method}
\end{figure*}

Fig. \ref{fig_method} conceptually illustrates the framework of analysis described above (panel \textit{a}), graphically showing the sPID approach and its decomposition terms (panel \textit{b}), as well as the estimation strategy \cite{bara2025partial} used in this work (panel \textit{c}). Mathematical details are provided in the following sections.

\subsection{PID framework}

Let us consider two source variables $X_1$ and $X_2$ and a target variable $Y$. The joint MI $I(Y;X_1,X_2)$ is defined as the information shared by the target and the pair of sources, intended as the average reduction in uncertainty about the outcomes of $Y$ when the outcomes of $X_1$ and $X_2$ are known. Exploiting the the framework of PID, this quantity can be expressed as the sum of unique contributions of the two sources to the target, i.e., $U(Y;X_1)$ and $U(Y;X_2)$, of their redundant contribution, i.e., $R(Y;X_1,X_2)$, and of their synergistic contribution $S(Y;X_1,X_2)$ \cite{williams2010nonnegative}, i.e.,
\begin{align}
    I(Y;X_1,X_2) &= U(Y;X_1) +U(Y;X_2)\nonumber \\
    &+R(Y;X_1,X_2)+S(Y;X_1,X_2).
    \label{estended_PID}
\end{align}
Considering the mathematical lattice structure representing the set-theoretic intersection of multiple variables (Fig. \ref{fig_method} (\textit{b})), the decomposition terms in Eq. \ref{estended_PID} can be seen as partial information (PI) quantities related to the information exclusively associated with each atom, represented by specific nodes of the lattice. Specifically, denoting the generic PI term as $I_\delta(Y;X_{\alpha})$ and considering the set of all possible subsets of sources $\mathcal{A}=\{\{1\},\{2\},\{1\}\{2\},\{12\}\}$, the joint MI can be compactly expressed as:
\begin{equation}
    I(Y;X_1,X_2) = \sum_{\alpha \in \mathcal{A}} I_{\delta}(Y;X_{\alpha}),
    \label{PID}
\end{equation}
with $I_\delta(Y;X_{\{1\}})= U(Y;X_1)$, $I_\delta(Y;X_{\{2\}})= U(Y;X_2)$, $I_\delta(Y;X_{\{1\}\{2\}}) = R(Y;X_1,X_2)$, and $I_\delta(Y;X_{\{12\}}) = S(Y;X_1,X_2)$.

The nodes in the redundancy lattice are ordered in such a way that atoms placed higher in the structure provide at least as much redundant information as those located lower, thus providing a structural ordering described by $\{1\}\{2\}\prec\{\{1\},\{2\}\}\prec\{12\}$, with $\prec$ indicating precedence.
Exploiting this ordering property, the PI of the generic atom $\alpha$ can be calculated iteratively from the related redundant information, denoted as $I_{\cap}(Y;X_{\alpha})$, i.e.,
\begin{equation}
    I_{\delta}(Y;X_{\alpha}) = I_{\cap}(Y;X_{\alpha})-\sum_{\beta\prec\alpha}I_{\delta}(Y;X_{\beta}), 
\end{equation}
where $\beta$ represents the atoms below $\alpha$ in the lattice structure and the term $I_{\cap}(Y;X_{\alpha})$ is the redundancy function defined according to the definition given by Williams and Beer \cite{williams2010nonnegative}:
\begin{equation}
    I_{\cap}(Y;X_{\alpha}) = \sum_{y\in A_Y}p(y)I_{\cap}(Y=y;X_{\alpha}),
\end{equation}
with $p(y)$ being the probability of each outcome of the target variable and $I_{\cap}(Y=y;X_{\alpha})$ the redundancy function related to the atom $\alpha$ and specifically associated with $Y=y$, which is defined as the minimum information that sources $\{X_s\}$ collected in the atom share with the specific outcome $y$ of the target variable, i.e., 
\begin{equation}
    I_{\cap}(Y=y;X_{\alpha})=\min_{\{X_s\}\in\alpha}I(Y=y;X_s).
\end{equation}
The PI term $I_{\delta}(Y;X_{\alpha})$ can therefore be expressed as the combination of PI terms associated with specific outcomes $y$ of the target variable $Y$, i.e., 
\begin{align}
    &I_{\delta}(Y;X_{\alpha}) = \sum_{y\in A_Y}p(y)I_{\delta}(Y=y;X_{\alpha})= \nonumber \\
    &=\sum_{y\in A_Y}p(y)\biggl(I_{\cap}(Y=y;X_{\alpha})-\sum_{\beta\prec\alpha}I_{\delta}(Y=y;X_{\beta})\biggr).
    \label{PI}
\end{align}
According to the formulation above, the central quantity to be determined is the information shared by the single or combined sources embedded in $X_s$ with the specific outcomes of the target, i.e., the MI $I(Y=y;X_s)$. 

\subsection{Estimation approach}

In this study, a nonlinear and nonparametric estimation approach, outlined in a previous study and well-suited for working with discrete target and continuous source variables, is employed \cite{bara2025partial}. Accordingly, the specific information shared by a continuous variable $X_s$, with outcomes $x_s \in D_{X_s}\subseteq \mathbb{R}^{n_s}$, and a specific outcome $y\in A_Y$ of the target can be expressed as the Kullback-Leibler divergence between the probabilities $p(x_s|y)$ and $p(x_s)$ \cite{bara2025partial}, i.e.,
\begin{equation}
    I(Y=y;X_s) = \int_{x_s\in D_{X_s}}p(x_s|y) \log\frac{p(x_s|y)}{p(x_s)}dx_s.
    \label{eq:spec_inf}
\end{equation}

The Kraskov-Stögbauer-Grassberger (KSG) nearest-neighbor MI estimator was used to compute this measure in a feasible way, allowing to take into account the computation of entropy terms on space of different dimensions, as well as the need for considering the entire samples or only those associated with specific outcomes of the target \cite{kraskov2004estimating}. Specifically, the distances $d_y$ to the $k^{th}$ neighbor are evaluated for the realizations of the joint sources associated to a specific outcome $y$ of the target variable, i.e., $\{(x_1,x_2)^y\}$, and then the specific information measures relevant to the whole source space $[X_1,X_2]$ and to the single source $X_n$, with $n\in\{1,2\}$, respectively computed as:
\begin{align}
    \hat{I}(Y=y;&X_1,X_2) = \psi(N)-\psi(N_y)+\psi(k) \nonumber\\
    &-\frac{1}{N_y}\sum_{\{(x_1,x_2)^y\}}  \psi(m_{X_1X_2}+1), \label{specMI_1}\\
    \hat{I}(Y=y;&X_n) = \psi(N)-\psi(N_y) \nonumber\\
    &+\frac{1}{N_y}\sum_{\{x_n^y\}}  \big( \psi(m_{X_n}^y+1)-\psi(m_{X_n}+1) \big), \label{specMI_2}
\end{align}
where $\psi(\cdot)$ is the digamma function, $N$ the total number of samples, $N_y$ the number of samples associated to the outcome $y$, $m_{X_1X_2}$ the number of observations of $[X_i X_j]$ at distance smaller than $d_y/2$ from the samples $\{(x_1,x_2)\}$, and $m_{X_n}^y$ and $m_{X_n}$ the number of observations of $X_n$ counted at distance smaller than $d_y/2$ from $x_n$ when considering all samples, i.e., $\{x_n\}$, or only those associated to the outcome $y$ of the target, i.e., $\{x_n^y\}$. 

We refer the reader to Ref. [\citenum{bara2025partial}] for further details on the estimation approach and to [\hyperlink{https://github.com/ChiaraBara/mfPID_toolbox}{GitHub}] for the freely downloadable MATLAB code.

\subsection{Statistical significance assessment}

The statistical significance of each of the PID decomposition terms, thus of the unique, redundant, and synergistic contributions of the sources to specific outcomes of the target, was assessed through a surrogate data analysis approach (see supplementary material of Ref. [\citenum{bara2025partial}]). 

The significance of the specific PI terms associated with each atom of the redundancy lattice is assessed by randomly shuffling the ordering of the samples of the target variable to destroy the correspondence between the target and the sources. The measure computed on the original data is then compared with a significance threshold derived from the measures evaluated on multiple surrogate data, obtained by repeating the shuffling procedure $N_s$ times. Specifically, considering the PI measure for each of the atoms of the lattice as the sum of $Q$ terms $I_{\delta}(Y=y;X_{\alpha})$ (see Eq. \ref{PI}), in accordance with the Bonferroni correction, the specific PI quantity is deemed as statistically significant if its value computed on the original data exceeds the $\left(100-\frac{0.05}{4Q}\right)^{th}$ percentile of the distribution obtained on surrogate data. Accordingly, if at least one of the $Q$ (or of the $4Q$) terms that are aggregated to derive each of the decomposition terms (or the joint MI) is deemed significant, the resulting measure is also significant.

\section{Application on simulated data} \label{Simul}

\subsection{Simulation setting and data generation} \label{sim_setting}
\begin{figure}[h!]
    \centering
    \includegraphics{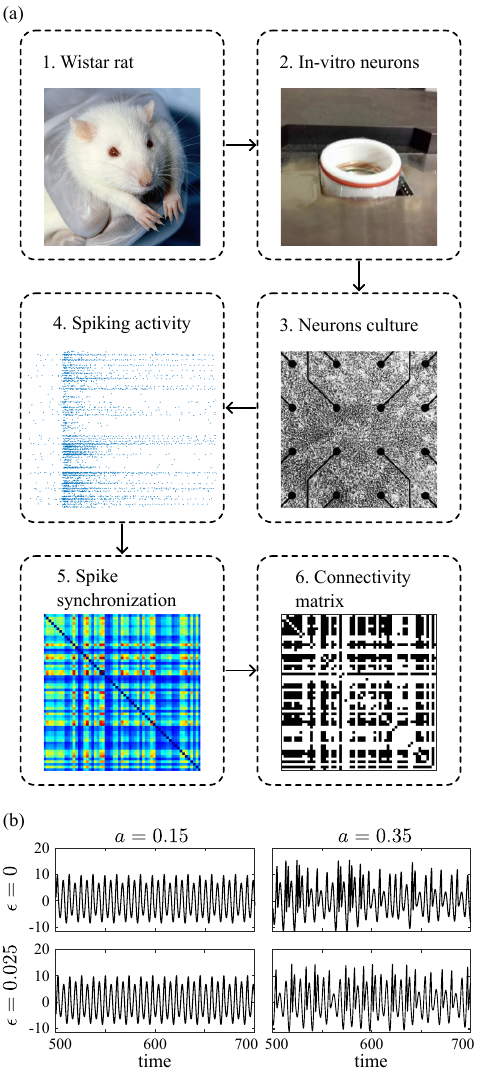}
    \caption{Graphical representation of the experimental study from the in-vitro cultured neurons of Wistar rats to the identification of the connectivity matrix representing brain-like network topology (panel \textit{a}) and 200-length windows of simulated $y$-variable time series for a representative R\"ossler unit (see Eq. \ref{Rossler_eq}) at varying the setting of the parameters $\epsilon$ and $a$ (panel \textit{b}). Part of the figure adapted from Ref. [\citenum{minati2025spontaneous}].}
    \label{fig1}
\end{figure}

The synchronized spontaneous bursting activity recorded from the cortex of embryonic Wistar rats was used to build a network topology that facilitates the numerical simulation of neural behaviors \cite{wagenaar2006extremely}.
Mechanically dissociated neurons were cultured on glass substrates equipped with multi-electrode arrays (MEAs) composed of fifty-nine electrodes to capture spiking activity across several neurons (Fig.~\ref{fig1} (\textit{a.1}, \textit{a.2})). Numerous cultures were obtained by plating approximately 50000 cells in a droplet that covers an area of $\sim5$ mm in diameter, with electrodes spaced by 200 $\mu$m, each capturing the dynamical activity of the ensemble of 100 to 1000 neurons (Fig.~\ref{fig1} (\textit{a.3})). Cultures exhibited diverse spiking behaviors and population-wide bursts of activity, with temporal dynamics evolving during maturation. For the subsequent analyses, data related to a maturation time of 24 days in vitro (DIV), which corresponds to the most advanced maturation state and a higher propensity to generate over-dispersed event distributions, were considered (Fig.~\ref{fig1} (\textit{a.4})). \\
The functional connectivity matrix was obtained by counting the recorded spiking activity across all electrodes in consecutive time windows of 25 ms and then thresholded and binarized to obtain the structural connectivity of the network (Fig.~\ref{fig1} (\textit{a.5}, \textit{a.6})). To retain only strong interactions and minimize noise artifacts, links associated with abnormally or highly connected nodes have been manually removed, while other links were added to ensure that each node had at least one link by randomly selecting the opposite end of the link with uniform probability \cite{minati2025spontaneous}. Accordingly, the 716 bivariate links initially counted were reduced to 676.
Further details of signal processing techniques used for spike and burst detection are reported in Ref. [\citenum{wagenaar2006extremely}].

Fifty-nine parametrically identical R\"{o}ssler units, i.e., three-dimensional systems of ordinary differential equations known for their paradigmatic chaotic behavior \cite{rossler1976chaotic, rossler1976equation}, were coupled according to the structural connectivity matrix obtained from the neural rats' spiking activity. 
For each unit $i=1,\dots,M$, the differential equations representing the dynamics of the oscillators are given by:
\begin{equation} \label{Rossler_eq}
    \begin{cases}
      \dfrac{\textrm{d}x_i}{\textrm{d}t}=-y_i-z_i\\
      \dfrac{\textrm{d}y_i}{\textrm{d}t}=x_i+ay_i+\epsilon A_{ij}(x_j-x_i)\\
      \dfrac{\textrm{d}z_i}{\textrm{d}t}=(x_i-c)z_i+b
    \end{cases},
\end{equation}
where $a$, $b$, and $c$ are the control parameters of the R\"{o}ssler systems, $\epsilon$ the coupling strength between pairs of oscillators along the $y$-variable, and $A_{ij}$ the generic element of the adjacency matrix $\boldsymbol{A}$ of dimension $M \times M$ which is equal to 1 if the unit $j$ drives the unit $i$ and 0 otherwise.
The bifurcation parameter $a$ that governs the chaotic behavior of all units was varied between 0.15 and 0.35 (for $a = 0.1$ the dynamics are periodic, while for $a \geq 0.2$ chaos ensues with increasingly rapid divergence and more developed folding, corresponding to the growth of branches on the Poincaré section). In contrast, the secondary parameters $b$ and $c$ were fixed at 0.2 and 5.7, respectively. The parameter $\epsilon$ governing the coupling strength ranged from 0 to 0.025. Both $a$ and $\epsilon$ can assume 48 values uniformly distributed across the indicated ranges.\\
All oscillator parameters were set identically in the nodes of the network and in the runs, while the initial conditions were randomized, setting $y_i(0) = z_i(0) = 0$ and $x_i(0) \in [-1, 1]$. The simulation was run up to 5500 time units, discarding the first 500 to ensure initial transient stabilization. The system was integrated with a time step of 0.1 using an explicit Runge–Kutta (4,5) formula, i.e., the Dormand-Prince pair. For both minimum and maximum levels of the parameters $a$ and $\epsilon$, the $y$-variable for a representative R\"{o}ssler unit is shown in Fig.~\ref{fig1} (\textit{b}) with a length of 200 time units.

The same simulation study was conducted in Ref. [\citenum{minati2025spontaneous}], where the system under investigation was characterized in detail in terms of phase synchronization and structural properties, as well as by examining the dynamic complexity of the units and their coupling and causality mechanisms for specific parameter setting configurations.

\subsection{Data and statistical analysis} \label{analysis_simul}
The framework of PID was applied on the numerically simulated R\"ossler network by considering the $y$-variables of pairs of oscillators within the network as sources, and the discrete network-state label obtained as a combination of the parameters $\epsilon$ and $a$ as target variable.

A total of $Q=36$ possible outcomes are considered for the target variable, corresponding to the combination of quantized levels of the two setting parameters across 6 levels for each, ranging from uncoupled synchronous to strongly coupled chaotic oscillators. Specifically, $N_r = 100$ realizations descriptive of the behavior of the network were obtained by randomly extracting time-alienated samples from the $y$-variable generated by all R\"ossler oscillators in the simulated network considering each possible combination of the forty-eight values of the two setting parameters $\epsilon$ and $a$, obtaining for each realization $N_s = 64$ samples that represent each of the $Q$ states of the system. This procedure ensured that the same probability of system states occurred within and among realizations. Consequently, the analysis was performed by working with realizations with a length of 2304 samples (equals to $N_s \times Q$).

For each pair of oscillators in the network, the joint mutual information between the discrete network-state label and the source variables was evaluated, together with its unique, redundant, and synergistic contributions assessed for each specific state of the network system.
All these terms were estimated as described in Sect.~\ref{method}, fixing a number of neighbors $k=5$; the statistical significance of each term was assessed by generating fifty surrogate data points.\\
Once all terms of the decomposition were computed, for each pair of sources, the unique contributions of the two oscillators were summed to obtain the combined unique information from both sources, thereby yielding the exclusive contribution of the pair. This combined contribution was deemed as significant if at least one of the two unique terms was significant. Moreover, the net redundancy-synergy balance was obtained as the difference between the redundant and synergistic terms. 

The joint mutual information computed for each pair of oscillators was related to the system's structure, thereby distinguishing between physically connected and non-connected oscillators. The same analysis was applied to the unique, synergistic, and redundant contributions, as well as to the net balance. For all these terms, the distributions of the average values on the one-hundred realizations were compared using an unpaired Student's \textit{t}-test with significance threshold of 0.05.
Moreover, another analysis was performed considering for each oscillator the average values of the information shared, of the combined unique contribution, and of the net balance obtained considering all its possible combinations with the other oscillators within the network, thus obtaining for each of these quantities a distribution on the one-hundred realizations of the system. The relation between these distributions and the degree of connectivity of the oscillators within the network, i.e., the number of physical connections of each node, was investigated via the mutual information (nearest-neighbor estimator, $k=5$) and statistically assessed by random shuffling the connection numerosity. Finally, the statistical significance of the redundancy-synergy net balance distributions was tested against zero by using a one-sample Student's \textit{t}-test. The significance level was set to $0.05$ for all statistical tests.

\subsection{Results}
\begin{figure*}[t!]
    \centering \includegraphics{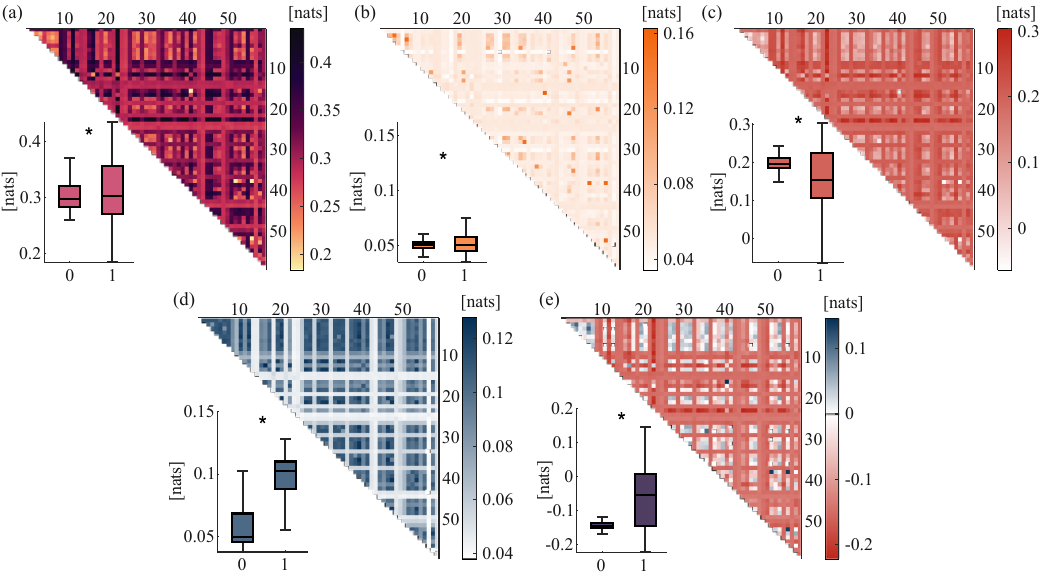}
    \caption{High-order interactions between pairwise oscillator activity and system state for the network of coupled R\"ossler systems. Upper triangular heatmaps depict the average values across the realizations of the information shared by each pair of oscillators and the state of the system (panel \textit{a}), as well as their unique (panel \textit{b}), synergistic (panel \textit{c}) and redundant (panel \textit{d}) contributions; the balance between the redundant and synergistic decomposition terms with positive values for net redundancy and negative values for net synergy is also depicted (panel \textit{e}). In each panel, boxplots depict the distribution of the average across realizations of the measures obtained for pairs of oscillators that are physically connected (1) or not (0). The line in the middle of the box represents the median of the distribution, while the bottom and upper limits of the box represent the 25$^{th}$ and the 75$^{th}$ percentile, respectively; the whiskers extend till the minimum and the maximum values of the distribution that are not considered as outliers. Statistically significant differences for each measure: $p<0.05$, 0 vs. 1, Student's \textit{t}-test.}
    \label{fig2}
\end{figure*}
Fig.~\ref{fig2} shows high values of joint MI for almost all pairs of oscillators within the system and its state (average significance of 100$\%$) (panel \textit{a}). The unique contribution is relatively low for all pairs of units (average significance of 97\%) (panel \textit{b}), while the synergistic and redundant contributions are higher (average significance of 100\%) for almost all pairs of units, indicating their co-existence also when considering the same pairs of oscillators (panels \textit{c}-\textit{d}). In addition, the measure of balance reveals a prevalence of synergy (panel \textit{e}). 

Considering for each oscillator the average synergistic and redundant information that it shares with all the other oscillators in the network, the net balance is significantly synergistic for all the oscillators.
The significantly slightly higher values of joint MI for pairs of oscillators physically connected are mainly related to the information redundantly shared by these oscillators, while unique and synergistic information contents are related mainly to not connected oscillators (panels \textit{b}-\textit{c}). 
Moreover, the correlation between the averaged PID terms for each unit and the number of its structural connections within the network reveals significantly higher values for the net redundancy-synergy balance (0.4644 nats) compared to the overall and uniquely shared information (0.1278 and 0.0853 nats, respectively).  

\begin{figure*}
    \centering
    \includegraphics{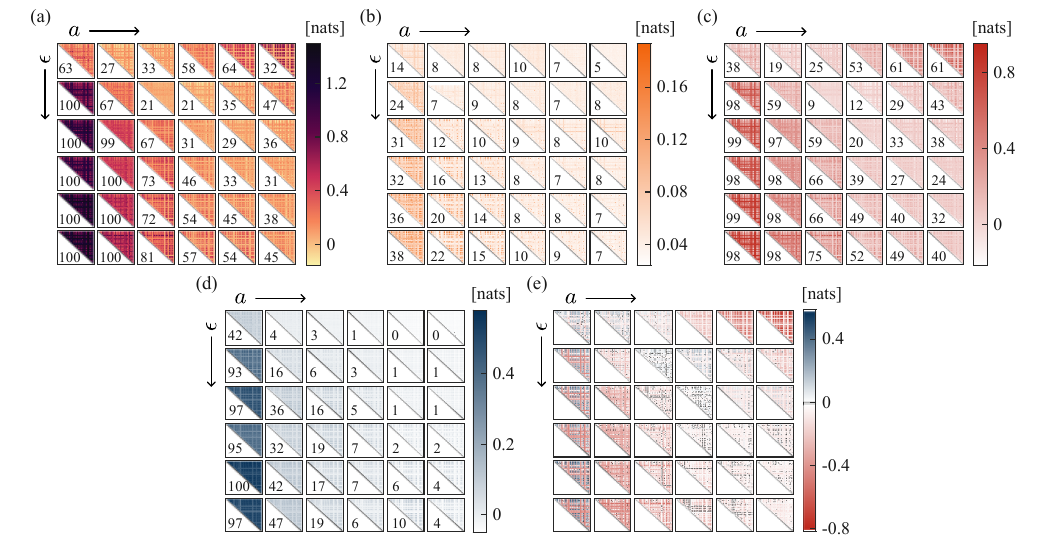}
    \caption{State-specific high-order interactions on simulated data. Upper triangular heatmaps depict the average values across the realizations of the information shared by each pair of oscillators and each specific state of the system (panel \textit{a}), i.e., by varying coupling strength $\epsilon$ in row and the bifurcation parameter $a$ in column, as well as their unique (panel \textit{b}), synergistic (panel \textit{c}), and redundant (panel \textit{d}) contributions; the balance between the redundant and synergistic decomposition terms is also depicted (panel \textit{e}). The percentage of realizations of the simulation for which the measure was deemed as significant (average over all pairs) is reported in the lower triangular heatmaps for all the possible thirty-six states of the system.}
    \label{fig3}
\end{figure*}

Fig.~\ref{fig3} shows how the information shared by pairs of units and the state of the system increases as $\epsilon$ increases and $a$ decreases, even with slightly high information shared for low values of $\epsilon$ and high values of $a$ (panel \textit{a}). The same trend is observed for the synergistic contribution (panel \textit{d}), while redundant and unique contributions only increase as $\epsilon$ increases and $a$ decreases (panels \textit{b}, \textit{d}). The trends of these quantities are also reflected in the results of the surrogate data analysis, which evidence increasing significance as the terms values increase. These results evidence again the simultaneous presence of synergy and redundancy, especially for high $\epsilon$ and low $a$, where the net balance of the overall system is quite balanced evidencing some pairs of oscillators presenting a prevalence of synergistic interplay and others with a prevalence of redundant interplay (panel \textit{e}). On the other hand, a net synergy appears significantly for low $\epsilon$ and high $a$ and a net redundancy for low $\epsilon$ and low $a$. 

\subsection{Commentary}
The high MI observed in the simulated system between the state of the network and pairs of $y$-variables referred to couples of R\"ossler oscillators, as well as for their synergistic and redundant contributions, demonstrates an inherent relationship between the behavior of the system and the dynamics of pairs of system units. The resulting complex patterns of interaction are characterized by a small to negligible unique contribution of individual links in determining the system's state, as well as by the coexistence of synergistic and redundant information shared between pairs of oscillators. Furthermore, the significant presence of both redundant and synergistic contributions, in addition to their modulation with the parameters that determine the behavior of the system and an overall prevalence of synergy, is indicative of the complexity and non-trivial functioning that characterize chaotic system oscillators, even in the presence of simple dyadic connections among the units.

In this system, redundant information exchange is likely due to the identical characterization of the multiple oscillators and their connections within the network. The existence of structural connections among the network's nodes has been shown to result in the synchronization of the oscillators: their time-aligned dynamics lead to the coordination of subparts within the network, thereby yielding redundant behavior of the involved units \cite{dorfler2013synchronization}. This finding is corroborated by the observation of considerably elevated redundancy values for physically connected oscillators, as well as by the increase in coupling strength within the network, which is known to exhibit strong synchronization.
However, a significant level of redundancy is also observed in disconnected oscillators, likely due to diffusive connections that facilitate synchronization among oscillators, even when they are not physically linked. Furthermore, it is crucial to emphasize the role of the bifurcation parameter $a$ in influencing redundant interaction among oscillators. Indeed, as demonstrated in \cite{minati2025spontaneous}, the oscillators within the network loose synchrony at high values of the bifurcation parameter, especially at low coupling strength, thereby weakening the redundant interactions.

Nevertheless, as observed, interactions among multiple chaotic oscillators lead to coordinated, structured dynamics, resulting in the emergence of strong synergistic behavior. Quite interestingly, the dominance of synergy over redundancy is emphasized when considering pairs of oscillators that are not structurally connected: this can be interpreted as an indication of high-order behavior that arises in the absence of high-order mechanisms, confirming the distinction between these two concepts \cite{rosas2022disentangling}. The increasing synergistic interplay with the coupling strength at low values of $a$ can be related to the aforementioned diffusive connection, which, together with overlapping information, also introduces additional information that leads to emergent behaviors. On the other hand, the increase in synergy with the bifurcation parameter at low coupling strength is reasonably associated with the introduction of novel interaction patterns due to chaoticity. 

\section{Application to experimental data}
\subsection{Circuit design and experimental measurements}

The biological phenomenon described in Sect.~\ref{sim_setting} also inspired a circuit-based experimental framework fully characterized in Ref. [\citenum{minati2025spontaneous}]. In this case, the connectivity matrix in Fig. \ref{fig1} (\textit{a.6}) was used as a blueprint to a physically wire large-scale electronic network based on Minati-Frasca electronic chaotic oscillators \cite{minati2017atypical}, i.e., atypical transistor-based chaotic oscillators that are known to provide rich, nonlinear, and chaotic dynamics, resembling those of biological neurons, especially in the ability to give rise to non-trivial synchronization patterns \cite{minati2024chaotic}. 

\begin{figure}
    \centering
    \includegraphics{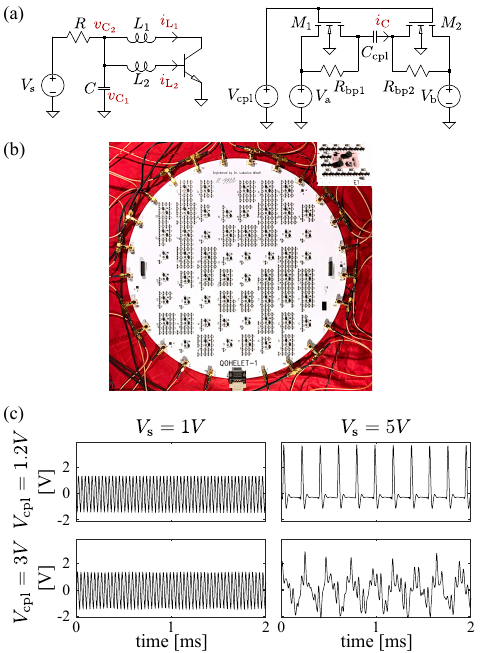}
    \caption{Single (left) and coupled (right) Minati-Frasca chaotic oscillator (panel \textit{a}), prototype of the circuit board (panel \textit{b}), and representative experimentally recorded time series at varying the setting parameters $V_{\textrm{cpl}}$ and $V_\textrm{s}$ (panel \textit{c}). Part of the figure adapted from Ref. [\citenum{minati2025spontaneous}].}
    \label{fig4}
\end{figure}

The Minati-Frasca oscillator depicted in Fig.~\ref{fig4} (\textit{a}, left) consists of a bipolar junction transistor of NPN-type in a grounded-emitter configuration, with its base and collector connected through two separated inductors ($L_1$ and $L_2$) and a resistor ($R$) in series with DC voltage source ($V_\textrm{s}$), and with a grounded capacitor ($C$). A system of equations describing the behavior of this circuit can be obtained in terms of the voltages at the capacitor and at the fictive capacitor between the collector and the emitter, $v_{\textrm{C}_1}$ and $v_{\textrm{C}_2}$, and of the currents through the two inductors, $i_{\textrm{L}_1}$ and $i_{\textrm{L}_2}$, i.e.,
\begin{equation}
    \begin{cases}
      \dfrac{\textrm{d}v_{\textrm{C}_1}}{\textrm{d}t}=\dfrac{V_\textrm{s}-v_{\textrm{C}_1}}{RC}-\dfrac{i_{\textrm{L}_1}+i_{\textrm{L}_2}}{C}\\
      \dfrac{\textrm{d}v_{\textrm{C}_2}}{\textrm{d}t}=\dfrac{i_{\textrm{L}_2}-i_\textrm{T}}{C_2}\\
      \dfrac{\textrm{d}i_{\textrm{L}_1}}{\textrm{d}t}=\dfrac{v_{\textrm{C}_1}-V_\textrm{th}}{L_1}\\
      \dfrac{\textrm{d}i_{\textrm{L}_2}}{\textrm{d}t}=\dfrac{v_{\textrm{C}_1}-v_{\textrm{C}_2}}{L_2}
    \end{cases}.
\end{equation}
The transistor base-emitter threshold voltage $V_\textrm{th}$ is typically fixed at 0.6 V, while the current $i_\textrm{T}$ measured at the fictive capacitor depends on the current of the base inductor $i_{\textrm{L}_1}$ and on the voltage at the collector $v_{\textrm{C}_2}$ by the function $i_\textrm{T} = \beta \gamma(i_{\textrm{L}_1})\tanh{(v_{\textrm{C}_2}/2V_\textrm{th})}$, with $\beta$ the current gain of the transistor and $\gamma(\cdot)$ the ramp function. Specific parameter conditions for which the oscillators do not assume neither the configuration of a relaxation-type oscillator, consisting of $R$, $L_2$, $i_\textrm{T}$ and $C_2$, nor a damped resonator oscillator, comprising of $R$, $C_1$ and $L_1$, lead to the emergence of chaotic dynamics. \\
The oscillators were connected through a voltage-controlled coupling circuit comprising a couple of MOSFET transistors, $M_1$ and $M_2$, biased by two base resistors, $R_{\textrm{bp1}}$ and $R_{\textrm{bp2}}$, and symmetrically controlling the current flow $i_C$ through a central coupling capacitor $C_{\textrm{cpl}}$. The coupling strength can be controlled by the external voltage $V_{\textrm{cpl}}$ (Fig.~\ref{fig4} (\textit{a}, right)).

The system described above was physically realized by integrating 59 oscillators on a round 4-layer circuit board; a prototype of the board is shown in Fig. \ref{fig4} (\textit{b}). Each Minati-Frasca oscillator consists of a single BCW66H bipolar transistor. Moreover, a capacitor of $C=$22 nF, a resistance of $R=$470 $\Omega$, and inductors of $L_1=$4.7 mH and $L_2=$10 mH (of types SRR6603-472ML and SRR6603-103ML, respectively) were employed.
The coupling circuits consist of resistive stages implemented with back-to-back N-MOSFETs (NX7002AKA) ensuring symmetric conductance; a $C_\textrm{cpl}=$1 $\mu$F AC-coupling capacitor and $R_\textrm{bp1}=R_\textrm{bp2}=$1 k$\Omega$ series resistor reference to a common VCPL rail are incorporated at each coupling stages allowing for a global adjustment of the coupling gain. 
A dual-output programmable power supply over IEEE-488 allows to systematically investigate the behavior of the realized board at varying the voltage source $V_\textrm{s}$, powering the fifty-nine oscillators across the eight distribution buses, and the external voltage $V_{\textrm{cpl}}$, controlling the overall coupling strength across the oscillator through the eight coupling buses.

The experiment was performed by varying the voltage source $V_\textrm{s}$ between 1 and 5 V and the external voltage $V_{\textrm{cpl}}$ in the range of 1.2-3 V. Specifically, each of the voltage quantities was varied considering 48 uniform levels in the range to respectively tune the chaotic behavior of all oscillators and the coupling strength in the overall circuit. For each acquisition stage, the output voltages of the 59 oscillators were simultaneously digitized at 12-bits with a sampling frequency of $f_\textrm{s}$=250 kHz to have recordings of 4 ms each. Time series windows lasting 2 ms are shown in Fig. \ref{fig4} (\textit{c}) for maximum and minimum values of both $V_\textrm{s}$ and $V_\textrm{cpl}$.

We refer the reader to Ref. [\citenum{minati2025spontaneous}] for further details on the experimental framework, circuit construction, and data acquisition.

\subsection{Data and statistical analysis}
Data preprocessing and analysis described in Sect.~\ref{analysis_simul} for the numerically simulated data was similarly performed for the experimental data acquired from the electronic oscillators circuit. Specifically, the sPID framework was employed to investigate the correlation between each pair of signals generated by the transistor-based oscillators within the circuit, intended as sources, and the discrete network-state label informative of the source and external voltages $V_\textrm{s}$ and $V_{\textrm{cpl}}$, taken as target variable representative on the chaoticity and coupling strength within the circuit.

Specifically, $N_r = 100$ realizations of the source variables describing circuit behavior were obtained by randomly extracting, for each value of the voltage source $V_\textrm{s}$ and of the external voltage $V_{\textrm{cpl}}$, the samples associated to each of the oscillator dynamics, thus obtaining 2304 samples characterizing the behavior of the 59 oscillators. Similarly to the application on simulated data, $Q=36$ possible outcomes were considered for the target variable, representing the coupling-chaos state of the circuit after quantizing the values of both $V_\textrm{s}$ and $V_{\textrm{cpl}}$ in 6 levels, thus obtaining $N_s=64$ samples representative of each of the $Q$ states of the system.\\
PID terms decomposing the joint mutual information between the coupling-chaos state of the system and the dynamics of each pair of oscillators were obtained by applying the same approach used for the simulated data. Moreover, the same statistical analyses were performed as to compare the results obtained on numerical and experimental settings.

\subsection{Results}
\begin{figure*}
    \centering
    \includegraphics{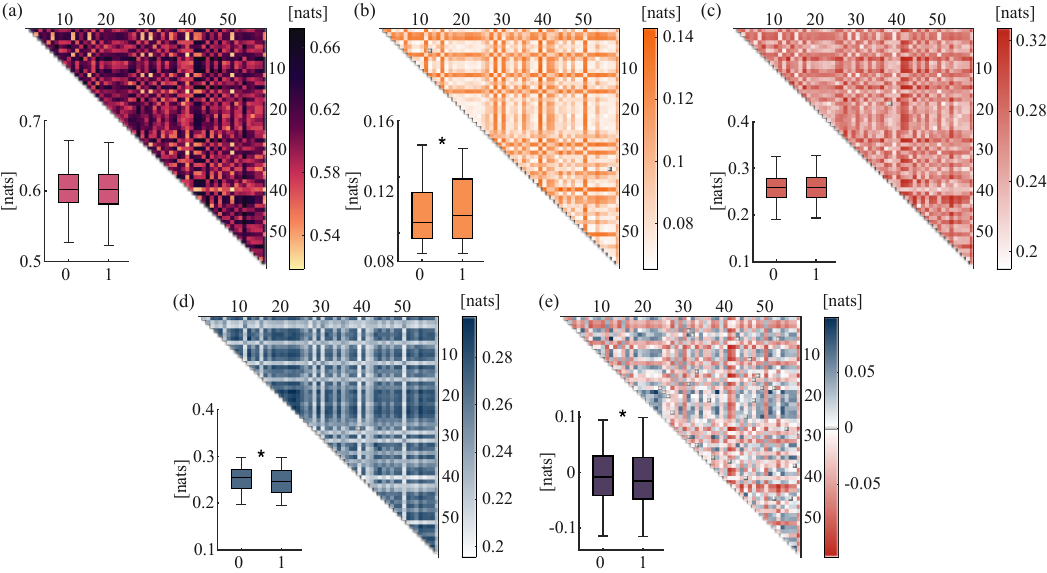}
    \caption{High-order interactions between pairwise oscillator activity and system state for the network of Minati-Frasca oscillators. Upper triangular heatmaps depict the average values across the realizations of the information shared by each pair of oscillators and the state of the system (panel \textit{a}), as well as their unique (panel \textit{b}), synergistic (panel \textit{c}) and redundant (panel \textit{d}) contributions; the balance between the redundant and synergistic decomposition terms with positive values for net redundancy and negative values for net synergy is also depicted (panel \textit{e}). In each panel, boxplots depict the distribution of the average across realizations of the measures obtained for pairs of oscillators which are physically connected (1) or not (0). The line in the middle of the box represents the median of the distribution, while the bottom and upper limits of the box represent the 25$^{th}$ and the 75$^{th}$ percentile, respectively; the whiskers extend till the minimum and the maximum values of the distribution that are not considered as outliers. Statistically significant differences for each measure: $p<0.05$, 0 vs. 1, unpaired Student's \textit{t}-test.}
    \label{fig5}
\end{figure*}

Fig.~\ref{fig5} shows high values of the information shared by all pairs of oscillators within the circuit and its state (average significance level of 100$\%$) (panel \textit{a}). These are related to overall high values of unique, synergistic, and redundant contributions (average significance level of 100$\%$) (panels \textit{b}-\textit{d}). As regards the net balance term, positive and negative values are distributed quite homogeneously among pairs of oscillators, with a slight prevalence of synergy in the whole (panel \textit{e}). Considering for each oscillator the average synergistic and redundant information shared with all the other oscillators in the network, the net balance is synergistic for 39 (out of 59) oscillators; in addition, while net synergy is always significant if tested against zero, the same does not hold for net redundancy (only 22 oscillators present significant net redundancy). These results suggest that synergy prevails over redundancy also considering signals produced by oscillators physically realized, even if such a prevalence is less marked than in the case of simulated oscillators presented in Sect. \ref{Simul}.

Even if no statistically significant differences are found for the joint MI when comparing oscillators which are physically connected or not, a slight increase of their unique contribution and a slight decrease of their redundant contribution (also reflected in terms of net balance), both statistically significant, are observed for physically connected oscillators (panels \textit{b}, \textit{d}, \textit{e}). 
Moreover, the correlation between the average PID terms related to each node of the network and the number of structural connections it forms shows higher values for the net balance measure (0.4028 nats vs. 0.1018 nats and 0.1637 nats of the overall and unique information, respectively).  

\begin{figure*}
    \centering
    \includegraphics{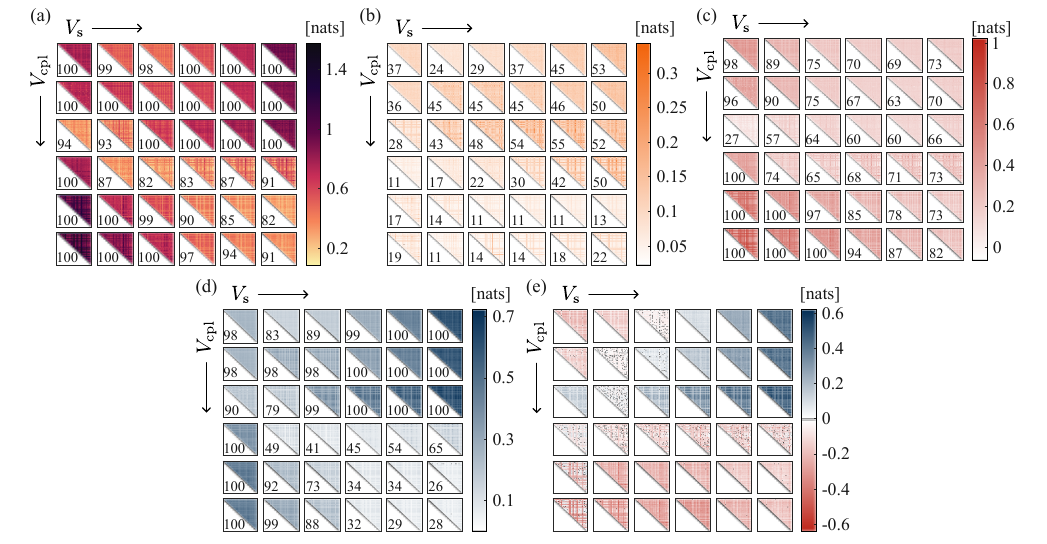}
    \caption{State-specific high-order results on experimental data. Upper triangular heatmaps depict the average values across the realizations of the information shared by each pair of oscillators and each specific state of the system (panel \textit{a}), i.e., by varying the external voltage $V_\textrm{{cpl}}$ in row and the voltage source $V_\textrm{s}$ in column, as well as their unique (panel \textit{b}), synergistic (panel \textit{c}) and redundant (panel \textit{d}) contributions; the balance between the redundant and synergistic decomposition terms is also depicted (panel \textit{e}). The average significance of all measures on the whole network is reported in the lower triangular heatmaps for all the possible thirty-six states of the circuit.}
    \label{fig6}
\end{figure*}

Fig.~\ref{fig6} shows higher average value of MI computed across pairs of source signals for the low values of $V_s$ and the high values of $V_{\textrm{cpl}}$ (panel \textit{a}), which cover also the states characterized by the strongest synergistic (panel \textit{c}) and redundant (panel \textit{d}) behavior. Moreover, high significant values of shared information are achieved also for low $V_{\textrm{cpl}}$ and high $V_\textrm{s}$, mainly related to the presence of high redundant information (panel \textit{d}).
These trends are also visible in terms of the net balance between redundancy and synergy, where the state associated with the lowest $V_\textrm{s}$ and higher $V_{\textrm{cpl}}$ present the co-existence of both redundant and synergistic behavior, while states related to higher synergy and redundancy are those associated to medium $V_{\textrm{cpl}}$/high $V_\textrm{s}$ and high $V_{\textrm{cpl}}$/medium $V_\textrm{s}$, respectively (panel \textit{e}).
As regards the unique contribution, it is overall lower than the redundant and/or synergistic contribution, but presents higher significant values for medium values of $V_{\textrm{cpl}}$ and high values of $V_\textrm{s}$ (panel \textit{b}). 

\subsection{Commentary}
The explored experimental configuration was developed as a parallel of the numerically simulated R\"ossler network through the use of transistor-based oscillators designed to replicate the nonlinear dynamics of neurons and the adjustable coupling employed to investigate diverse synchronization regimes. The same analysis performed on numerically simulated and experimentally generated data yields comparable results, e.g., the coexistence of synergistic and redundant contributions of pairs of units in determining the state of the system, especially at high coupling strength and low chaoticity. Nevertheless, there are also disagreements regarding the units' interplay in high-chaos and medium/low-coupling contexts, as well as the trends in the information terms when considering the presence or absence of a structural connection between them.

This behavioral heterogeneity exhibited by the two analyzed network systems in terms of trends of the PID terms with both the state of the network and the physical connection of the oscillators is first indicative that the redundant and synergistic sharing of information among oscillators is not attributable to the structural connectivity scheme, as this remains consistent in both systems. The primary distinction between the two systems concerns the ideality of the simulated network and the realism of the physical system. While the former is composed of parametrically identical units, the latter comprises several electronic components that, even if the oscillators are structurally identical, can contribute to the 50\% difference observed when comparing the oscillators in the network. More specifically, the typical tolerance of electronic component parameters is 1\% for resistors, 5\% for capacitors, 20\% for inductors, and up to 50\% for transistors. Furthermore, electronic noise and physical circuit imperfections affect the operation of the hardware system. These spurious terms may be attributed to the less-delineated patterns of interaction observed in the experimental setup compared to the simulated one, as well as to differences in physical connections and network state.

\section{Application to physiological data}

\subsection{EEG signals acquisition and pre-processing}

Electroencephalographic (EEG) signals obtained from 21 young, healthy subjects (10 males and 11 females, aged 22-39 years) with normal vision and no history of neurological or mental diseases were analyzed. 19 EEG channels were recorded using the 10-20 standard system for electrode positioning, with Fpz and Oz electrodes set as reference and ground electrodes, respectively (Micromed Brain Quick System, Treviso, Italy). The acquired data was then digitized with a sampling rate of $f_\textrm{s}$ = 256 Hz and 16-bit resolution.
The participants were instructed to lie down and assume a relaxed state in a dark room free from electrical and acoustic interference. Then, brain neural signals were acquired for 40 seconds with both eyes closed (EC) and eyes open (EO). Further details on data acquisition can be found in Ref. [\citenum{faes2016information}]; no human experiment was directly performed as a part of this study.

EEG signals were filtered through a zero phase-shift band-pass filter between 0.3 and 40 Hz, and, if necessary, any artifacts resulting from eye blinks, eye movements, and cardiac activity were removed using Independent Component Analysis (ICA). Subsequently, all EEG signals were re-referenced to a common average voltage and down-sampled to $f_\textrm{s}=$128 Hz. Finally, 8-second windows (corresponding to 1024 samples) were selected for each subject and condition through an iterative test that checked the restricted weak stationarity of the signals in mean and variance \cite{magagnin2011non}. 

\begin{figure}
    \centering
    \includegraphics{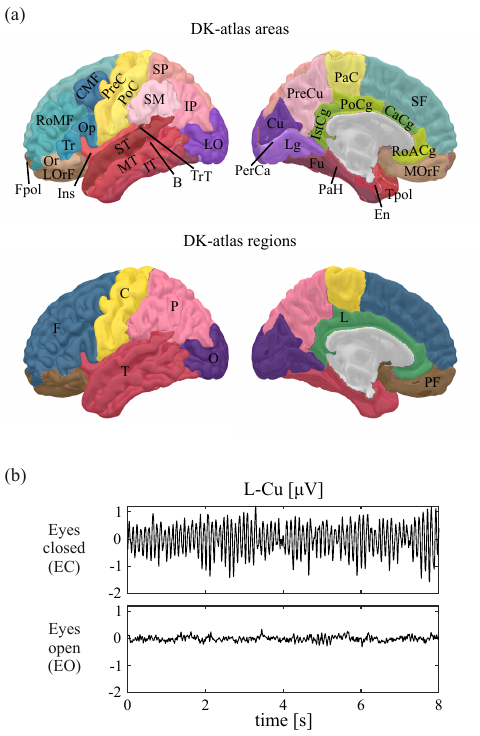}
    \caption{Graphical representation of brain areas according to DK atlas and their grouping in large regions (panel \textit{a}): in yellowish the central (C) cortex (pre central, PreC; post central, PoC; para central, PaC), in bluish the frontal (F) cortex (rostral middle frontal, RoMF; caudal middle frontal, CMF; pars opercularis, Op; pars triangularis, Tr; superior frontal, SF), in greenish the lateral (L) cortex (isthmus cingulate, IstCg; posterior cingulate, PoCg; caudal lanterior cingulate, CaCg; rostral anterior cingulate, RoACg), in purplish the occipital (O) cortex (lateral occipital, LO; cuneus, Cu; pericalcarine, PerCa; lingual, Lg), in pinkish the parietal (P) cortex (superior parietal, SP; supra marginal, SM; inferior parietal, IP; pre cuneus, PreCu), in brownish the prefrontal (PF) cortex (pars orbitalis, Or; lateral orbitofrontal, LOrF; frontal pole, Fpol; medial orbitofrontal, MOrF), and in reddish the temporal (T) cortex (insula, Ins; superior temporal, ST; middle temporal, MT; inferior temporal, IT; transverse temporal, TrT; bankssts, B; fusiform, Fu; para hippocampal, PaH; enterhinal, En; temporal pole, Tpol). 8-second windows of source data referred to the left Cuneus (L-Cu) area for a representative subject with both eyes closed (EC) and eyes open (EO) (panel \textit{b}).}
    \label{fig7}
\end{figure}

These signals were projected via spectral inverse operators to their cortical sources using the CiftiStorm pipeline (a MATLAB toolbox for spectral analysis and inverse solution of electroencephalographic and magnetoencephalographic data) to obtain the head model, source model, and lead field required for calculations of the inverse operators \cite{areces2024ciftistorm}. Specifically, the spectral analysis was performed via a \textit{k}-order orthogonal wavelet decomposition of the data, employing the maximal overlap discrete wavelet transform method (MODWT) \cite{percival2000wavelet}. Accordingly, the wavelet transforms are obtained for the vector functions composed of the EEG signals to have peaks around bands of frequency described by the rule $\big( \frac{f_\textrm{s}}{2^{k+1}},\frac{f_\textrm{s}}{2^k}\big]$. In our settings, a maximum wavelet order was set to $k=7$ to have waveforms peaking around the fundamental EEG frequency bands, i.e., gamma $\gamma$ (32-64 Hz) for $k=1$, beta $\beta$ (16-32 Hz) for $k=2$, alpha $\alpha$ (8-16 Hz) for $k=3$, theta $\theta$ (4-8 Hz) for $k=4$, high delta h-$\delta$ (2-4 Hz) for $k=5$, mid delta m-$\delta$ (1-2 Hz) for $k=6$, and low delta l-$\delta$ (0.5-1 Hz) for $k=7$. The inverse solution was then obtained via the Spectral Structured Sparse Bayesian Learning (SSSBL) method operating via an optimal quasilinear operator applied to the wavelet transform data \cite{paz2023minimizing}. This approach optimizes the operator to identify the source activity patterns produced by cortical oscillatory networks at each of the major frequency bands in resting-state EEG. Quasilinearity is a key quality of the SSSBL approach, which pursues strong regularization via nonlinear priors (specifically the Elastic Net Nuclear Norm), thereby controlling spatial distortions or leakage of the source time series \cite{gonzalez2018caulking}, while avoiding nonlinear distortions in both the time and frequency domains. Specifically, the Desikan-Killany (DK) atlas, consisting of 34 cortical regions for hemisphere (both left (L) and right (R)), was employed \cite{desikan2006automated}; a graphical representation of the DK areas and of their possible grouping in larger cortical regions is depicted in Fig. \ref{fig7} (\textit{a}).

\subsection{Data analysis}

The PID analysis was performed on source data to investigate the nature of the information carried by pairs of cortex signals about the experimental condition (eyes open vs. eyes closed).

The samples of the signals from different cortical areas were considered as realizations of the source variables, while the acquisition phase was treated as a binary output of the target variable ($Q=2$), which consisted of EO and EC states. Thus, $N_s = 1024$ samples for each of the $Q$ conditions were analyzed for each subject (for a total data length of 2048 samples) and the analysis was repeated for the 2278 combinations of the 68 identified brain areas considering both hemispheres. The PID method was applied by using a number of neighbors $k=5$. Moreover, the statistical significance of each term was assessed by generating 50 surrogate data by shuffling the samples of the target variable. 

As in both simulation and experimental setups, the unique contributions of pairs of areas were summed to obtain a combined unique information, and the redundancy-synergy balance was investigated by considering the difference between redundant and synergistic contributions from each pair of areas. Finally, to simplify the reading of the results, averaged values and significance for the larger cortical regions are obtained for each measure, considering the mean of the terms associated with pairs of cortical areas corresponding to those regions.

\subsection{Results}

\begin{figure*}
    \centering
    \includegraphics{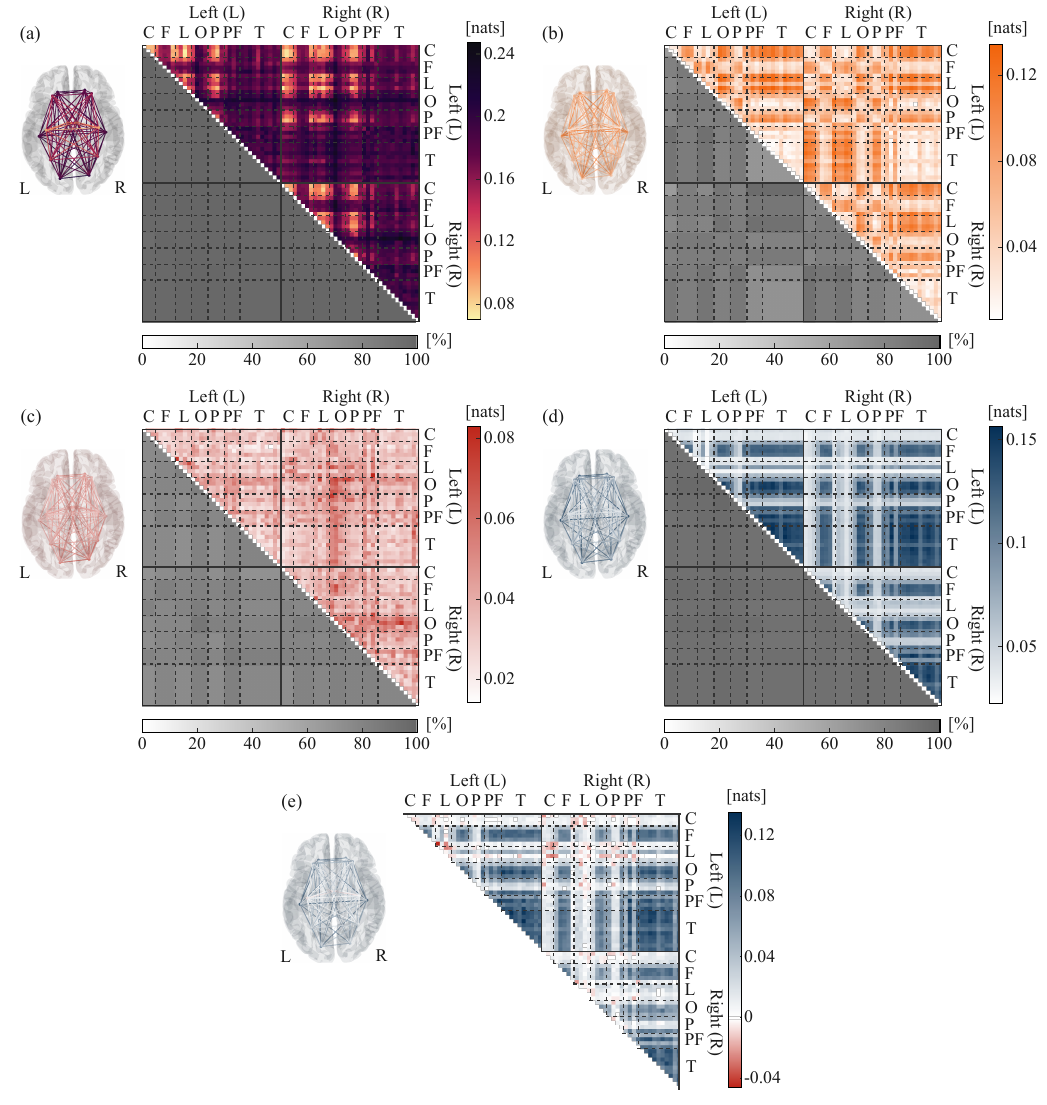}
    \caption{Heatmaps depicting in the upper triangular the average values across subjects of the information shared by each pair of cortical source and the acquisition phase  (panel \textit{a}), as well as their unique (panel \textit{b}), synergistic (panel \textit{c}) and redundant (panel \textit{d}) contributions, as well as the balance between the redundant and synergistic decomposition terms (panel \textit{e}), and in lower triangular the correspondent average significance evaluated within and between left (L) and right (R) cortical regions (central (C), frontal (F), lateral (L), occipital (O), parietal (P), prefrontal (PF), and temporal (T)).
    On the left, a 3D cortex representation considering the average values of the corresponding measure within (colored nodes in the centroid of each region) and between (colored links between pairs of centroids) cortical regions.}
    \label{fig8}
\end{figure*}

Fig.~\ref{fig8} shows high values of joint MI between the acquisition's state and pairs of signals of all cortical areas except central, lateral and parietal; 100\% significance is observed in the whole cortex (panel (\textit{a})). Besides a slight high synergistic contribution when considering at least one of the source variables from areas in the occipital regions (panel (\textit{c})), high joint MI is mainly related to the redundant contributions observed when jointly considering areas involved in frontal, prefrontal, occipital, and temporal regions (panel (\textit{d})), as well as to the unique contributions associated with the interactions of one of these regions with the other regions of the cortex, i.e., central, lateral, and parietal (panel (\textit{b})). As expected, for each term, high values are reflected in higher significance, with overall higher significance for the redundant contribution. The areas mainly presenting redundant contributions are also evidenced in terms of net balance, while the other present a quite balanced condition or a slight shift towards synergy (panel \textit{e}).

\begin{figure*}
    \centering
    \includegraphics{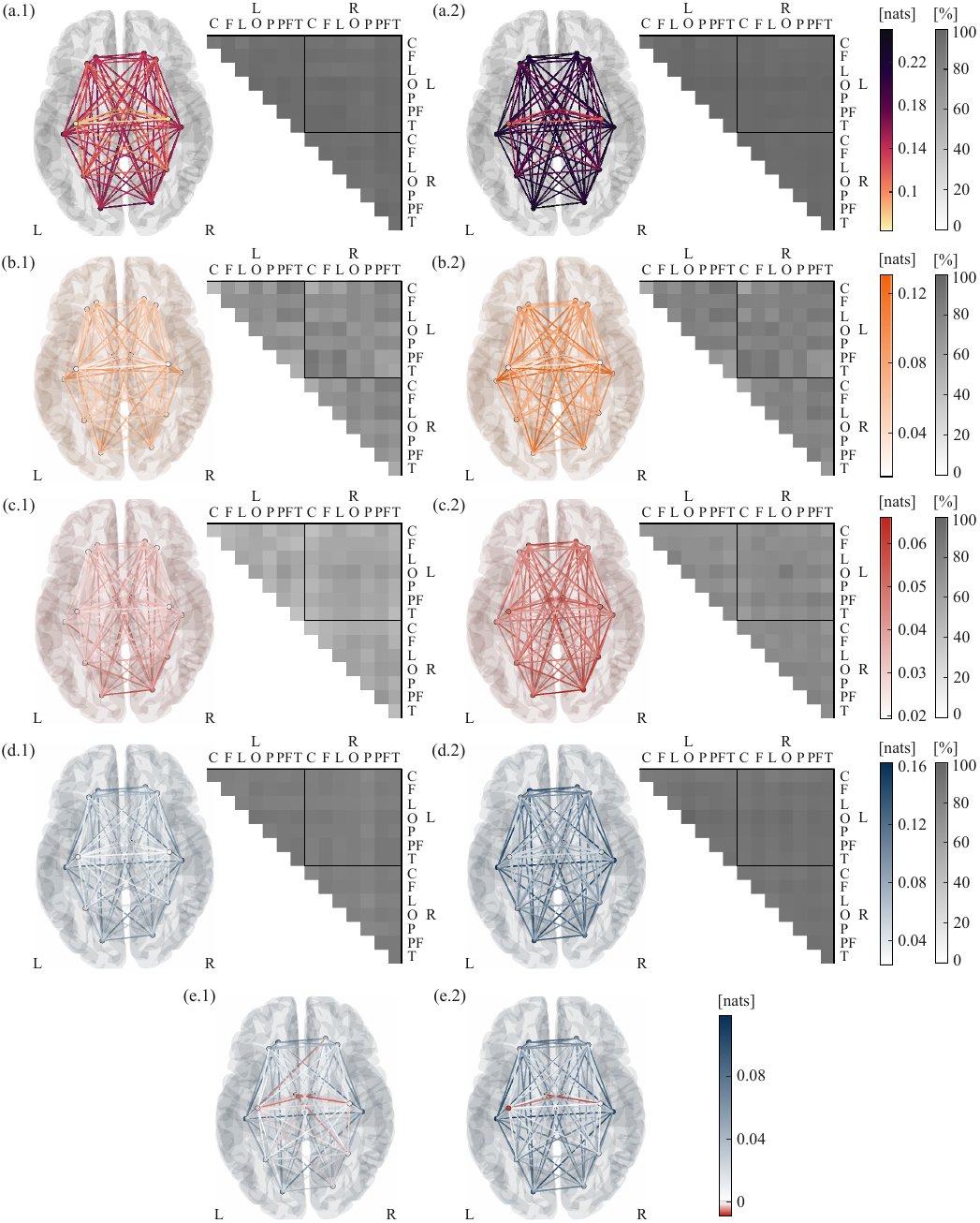}
    \caption{3D cortex representations of average values across subjects of the averaged measures within (colored nodes) and between (colored links) cortical regions. Specifically, values related to the information shared by each pair of regions and the acquisition phase (panels \textit{a.}), as well as their unique (panels \textit{b.}), synergistic (panels \textit{c.}) and redundant (panels \textit{d.}) contributions are shown for each specific phase of the acquisition protocol, i.e., eyes closed (panels \textit{.1}) and eyes open (panels \textit{.2}). Additionally, the balance between the redundant and synergistic decomposition terms is also depicted (panels \textit{e.}). 
    In each panel, the upper triangular heatmaps depict the average significance evaluated within and between left (L) and right (R) cortical regions (central (C), frontal (F), lateral (L), occipital (O), parietal (P), prefrontal (PF), and temporal (T)).}
    \label{fig9}
\end{figure*}

Fig.~\ref{fig9} confirms the involvement of the cortical regions already highlighted for the various information contributions, exhibiting higher values when subjects are in the EO condition (panels (\textit{.2})). This result is also reflected in an increase in the average statistical significance of each measure across the entire cortex from EC to EO, specifically from 94\% to 97\% for the joint MI, from 71\% to 80\% for the unique term, from 54\% to 74\% for the synergistic term, and from 83\% to 91\% for the redundant term.

\subsection{Commentary}
Investigations of cortical data reveal intricate patterns of information arising from the interplay between brain regions when eyes are opened or closed. The well-known involvement of the occipital cortex in early visual processing, thus its increased synchronized activity in the alpha band when the eyes are closed \cite{parks2013brain}, is supported by our results which highlight a high joint MI between cuneus (Cu), pericalcarine (PerCa) and lingual (Lg) areas, constituting the visual network (VN), and the other cortex regions with the co-presence of a high redundant and synergistic contributions. 
Nevertheless, engagement of other cortical regions, such as the frontal, prefrontal, and temporal regions, was observed. Specifically, dorsal attention network (DAN) comprising the frontal eyes field (FEF), associated with the activity of the middle frontal gyrus structural area in the frontal cortex region (corresponding to caudal middle frontal (CMF) and rostral middle frontal (RoMF) DK-atlas areas) \cite{blanke2000location}, has been observed to be involved in the voluntary shifts of visual attention, as well as in the eye movement control \cite{lynch2006cortico}, with a high alpha activity in the response preparation to a visual stimulus \cite{popov2017fef} and when eyes are closed \cite{marx2004eyes}. 
Similarly, although the temporal cortex is known to be primarily involved in higher-level visual functions (e.g., object recognition and feature integration) \cite{grill2004human}, its engagement during simple visual tasks involving repeated or familiar stimuli reveals a early recruitment and top-down facilitation mechanisms \cite{bar2006top, bell2009object} that, together with a hierarchical and parallel organization of the ventral visual stream which enables the rapid transmission of visual information from occipital areas to temporal regions \cite{felleman1991distributed,lamme2000distinct}, results in dynamics similar to the VN.
Our findings support these involvements through the presence of significant redundant information exchange within and between frontal, temporal, and occipital regions, suggesting highly correlated oscillatory dynamics in these regions. 

The results observed in the physiological data are in reasonably good agreement with those obtained in both simulated and experimental settings. Specifically, highly synchronized units (in this case, cortical regions) results in the coexistence of significant synergistic and redundant contributions -- look at synergistic interactions within and with the occipital region -- and non-ideal settings (so hardware and physiological ones) are characterized by the prevalence of a redundant exchange of information in determining the state of the system -- looks at the redundant interplay within and between occipital, frontal, parietal, and temporal cortical regions. Nevertheless, when working with physiological data, a more evident redundant behaviour is observed than in the electronic network, probably due to the different nature of the established interplays among units, which are mainly related to functional dependencies in the brain and to structural connections in the hardware system \cite{rosas2022disentangling}. The observation of prevailingly redundant information shared between pairs of source EEGs and the system state is consistent with the concept that redundancy is related to robustness, a notion enforced particularly in neuroscience applications of PID related to the study of human sources of visual information \cite{luppi2024information}.

Finally, the observed comparable trends in the PID terms within the cortex in EC and EO states are expected to result from consistent activation and deactivation of spatial patterns during the two phases of the protocol. However, the higher contributions observed in the EO condition may be attributed to the greater exchange of information involving the brain when eyes are open, due to the presence of external stimuli and wider oscillatory activity \cite{gotz2017influence, petro2022eyes}.

\section{Discussion and conclusions}

The present study investigates multiple-domain systems to elucidate non-trivial behaviors arising in network architectures. The extant literature supports the hypothesis that, in large systems, subparts cooperate to determine functional structures at higher levels of the network \cite{haken1977synergetics}. Relatedly, according to the theory of causal emergence, the state that characterizes the overall system can be seen as a supervenient observable feature \cite{rosas2019quantifying}. In recent years, these concepts have driven research into approaches that delve into these intricate interactions and distinguish the unique information contribution provided by individual subsystems, as well as the information they equally share, which is therefore redundant, or that arises from their interaction, which is therefore synergistic \cite{timme2014synergy}. Among the extensive range of approaches, the analytical framework of PID stands out thanks to its ability to differentiate between redundant and synergistic information content \cite{williams2010nonnegative}. 
In this work, PID was applied using a novel perspective reflecting the aforementioned idea that the state is a supervenient feature of the system. Specifically, while traditional PID problems follow a top-down approach in which the high-order interactions among triplets of nodes are investigated for a fixed state of the system \cite{sherrill2021partial, bara2025partial}, the proposed sPID framework employs a bottom-up approach in which the aim is to examine the state of the system as one of the variables of the analysis, which varies in accordance with the manner in which pairs of system's nodes interact.

Thus the application of sPID facilitates the identification of the relationship between signals generated by or acquired from pairs of nodes and the state of the system. This is achieved by examining the measure of joint MI, determining the type of information shared by these signals, and assessing the components of PID to determine the overall state of the system. Moreover, extending the analysis to all possible pairs of nodes in the network enables a thorough investigation of the system's behaviour. Specifically, this investigation provides insights into the unique contribution of individual nodes to the system's state, the redundant contribution of functionally segregated nodes that provide the same information, and the synergistic contribution of functionally integrated units that collectively provide new information to the system.
In this regard, it is important to emphasize that the focus of our approach is on the behavior of networked systems, rather than on the mechanisms underlying the networks \cite{rosas2022disentangling}. Consequently, the emphasis is placed on the functional dependencies between the signals attributed to the nodes, as opposed to the topological characterization of the network generally explored by means of simplicial complexes \cite{zhang2023higher, robiglio2025synergistic}.  
Furthermore, it is important to emphasise that the analysis performed using this approach is based on the static PID. Indeed, despite recent developments in dynamical approaches to account for the impact of temporal correlation on PID components \cite{faes2025partial}, the static approach is deemed more suitable for the purposes at hand. This is because its primary focus is not on investigating signal interactions, but instead on their correlation with the system's state, which can be considered a zero-lag interaction.

The present analysis, which was conducted on complex network systems composed of numerically simulated R\"ossler oscillators and physical Minati-Frasca chaotic oscillators, as well as of cortical areas, reveals both analogies and differences among the three systems. We found that pairs of oscillators, in both numerical and experimental contexts, exhibit both redundant and synergistic information when considering overall highly synchronized systems, i.e., when the simulation is set to high $\epsilon$ and low $a$ and the physical system is set to high $V_\textrm{cpl}$ and low $V_\textrm{s}$. A similar behavior was also observed in the brain network, with areas grouped into networks that appear to synchronize to the same oscillatory dynamics during the execution of visual tasks. 
These findings are in line with expectations: nodes that are synchronized have been shown to exhibit overlapped dynamics \cite{dorfler2013synchronization}, thus to share the same information in determining the state of the system, but in addition, they also contribute to finalizing and straightening the effect on the system's outline, revealing synergistic information. The synergistic contribution may also be partly related to the presence of complex dynamics of internal nodes, which are not entirely obscured by synchronization. As a consequence, the observed increase in the redundant information for connected oscillators in the simulated settings \cite{dorfler2013synchronization}, as well as the greater synergistic information observed in the neural network when considering the interplay of the occipital region with the other areas involved in visual tasks \cite{lynch2006cortico, blanke2000location}, are expected. 
Nevertheless, as evidenced by the net balance, the high significance of both contributions is clear in the ideal, completely controlled simulated system, which overall shows a prevalence of synergistic information. However, synergy decreases in the physically realized system, even if it remains prevalent, and even more so in the physiological setting, where redundancy is overall prevalent.
In addition to the differences related to the ideality of the simulated system and the realism of the other two systems, to which the higher redundant informative contributions can be referred to due to the presence of electronic tolerances in the oscillators' network and more generally to the possible external disturbances which similarly affect all the nodes of the real networks, the differences observed across the networks can also be attributed to the different nature of the connections and the scale of visualization of the phenomena in oscillators' and brain networks. Specifically, oscillator circuits are characterized by the structural connections between units, and the target variable describes the mesoscopic state of the system \cite{minati2015synchronization}. Conversely, neural networks deal with functional connections, and the target variable describes the macroscopic state of the system. This can lead to divergent interpretations of the information-sharing patterns. 

As previously mentioned, this analytical framework is highly applicable to the domain of network neuroscience, underpinning complex cerebral functions resulting from a balance between distributed integration and local specialization \cite{tononi1994measure, bassett2017network}. Indeed, as the coexistence of synergy and redundancy is considered a hallmark of healthy neural complexity \cite{luppi2022synergistic}, the balance between functional integration and functional segregation is regarded as an indicator of healthy brain \cite{tononi1994measure, sporns2004organization, zuberer2021integration} and a disruption of this equilibrium may be indicative of neuropathological states, such as Alzheimer's disease or schizophrenia \cite{dickerson2009large, fornito2015reconciling}. 
More specifically, it is expected to observe redundancy and synergy when looking respectively at the interactions within and between segregated specialized regions, as well as when considering the integration of different specialized areas.
In this view, this work aligns with a growing body of recent studies that investigate these mechanisms using information-theoretic metrics \cite{luppi2022synergistic, varley2023multivariate, pope2025time}.

Despite the relevance and robustness of the achieved results, further advancements are required to enhance our study. 
Firstly, to achieve a more comprehensive characterization of the intricate behavior exhibited by such network systems, it is necessary to move beyond pairwise interactions and investigate the dynamics of groups of oscillators or larger brain regions in a hierarchical manner. This multiscale PID approach promises to reveal how information integration unfolds hierarchically within the brain. Furthermore, it is pertinent to emphasize that the methodology employed is grounded in the definition of a redundancy measure, whose optimal formula remains a subject of debate in the extant literature. A plethora of measures have been defined as representative of redundant information, based on the numerous properties that are supposed to be accounted for by this term. However, there is no consensus \cite{bertschinger2012shared, ince2017measuring}. In this work, we applied the seminal definition of Williams and Beer, which, by considering the formulation proposed in Ref. [\citenum{bara2025partial}], allowed us to work with mixed discrete and continuous variables.
Furthermore, regarding the physiological application, even if a sophisticated approach was employed to reconstruct cortical signals from scalp traces, the reduced number of EEG channels must be considered a possible influencing factor on the neural application results. The phenomenon of signal leakage may yield estimates of the source that are not entirely related to the execution of the task, potentially leading to an overestimation of redundantly shared information attributable to methodological artefacts.

\bibliography{biblio.bib}

\end{document}